\documentclass[preprint]{cimento}

\usepackage{graphicx}  

\newcommand{\bra}[1]{\langle #1|} 
\newcommand{\ket}[1]{|#1\rangle}

\newcommand{\lt}{\left}
\newcommand{\rt}{\right}

\newcommand{\gev}{\,\mbox{GeV}}
\newcommand{\mev}{\,\mbox{MeV}}

\newcommand{\bbd}{$\mathrm{B_d}\!-\!\ov{\mathrm{B}}{}_\mathrm{d}\,$}

\newcommand{\bbq}{$\mathrm{B_q}\!-\!\ov{\mathrm{B}}{}_\mathrm{q}\,$}
\newcommand{\bbms}{$\mathrm{B_s}\!-\!\ov{\mathrm{B}}{}_\mathrm{s}\,$\ mixing}
\newcommand{\bbmd}{$\mathrm{B_d}\!-\!\ov{\mathrm{B}}{}_\mathrm{d}\,$\ mixing}
\newcommand{\bbmq}{$\mathrm{B_q}\!-\!\ov{\mathrm{B}}{}_\mathrm{q}\,$\ mixing}

\newcommand{\bbm}{$\mathrm{B}\!-\!\ov{\mathrm{B}}{}\,$\ mixing}
\newcommand{\kkm}{$\mathrm{K}\!-\!\ov{\mathrm{K}}{}\,$\ mixing}

\newcommand{\dm}{\ensuremath{\Delta m}}
\newcommand{\dg}{\ensuremath{\Delta \Gamma}}
\newcommand{\bea}{\begin{eqnarray}}
\newcommand{\eea}{\end{eqnarray}}

\newcommand{\ov}{\overline}
\newcommand{\imag}{\mathrm{Im}\,}

\newcommand{\eq}[1]{eq.~(\ref{#1})}

\newcommand{\fig}[1]{fig.~\ref{#1}}

\title{{\small TTP10-27 \hfill SFB/CPP-10-46}\\[5mm]
Flavour physics within and beyond the 
Standard Model$^2$}
\author{U.~Nierste\from{ins:x}}
\instlist{\inst{ins:x}Karlsruhe Institute of Technology\\ 
          Universit\"at Karlsruhe\\ 
          Institut f\"ur Theoretische Teilchenphysik\\
          76128 Karlsruhe, Germany}

\PACSes{12.60.Jv,13.20.He,12.10.-g}

\begin{document}
\maketitle\footnotetext[2]{Invited talk at \emph{Les Rencontres de Physique
     de la Vall\'ee d'Aoste}, La Thuile, Italy, Feb 28 -- Mar 6, 2010.}

\begin{abstract}
  I review recent progress in theoretical calculations related to the
  CKM unitarity triangle. After briefly discussing hints for new physics
  in \bbd\ and \bbms\ I present three topics of MSSM flavour physics:
  First I discuss new $\tan\beta$-enhanced radiative corrections to
  flavour-changing neutral current (FCNC) amplitudes which go beyond the
  familiar Higgs-mediated FCNC diagrams and may enhance the
  mixing-induced CP asymmetry in $B_d\to \phi K_S$. The second topic is
  a reappraisal of the idea that flavour violation originates from the
  soft supersymmetry-breaking terms. Finally I discuss how $\mu \to e
  \gamma$ can be used to constrain the flavour structure of the
  dimension-5 Yukawa interactions which appear in realistic grand
  unified theories.
\end{abstract}

\section{Introduction}
Flavour physics addresses the transitions between fermions of different
generations. Within the Standard Model these transitions originate from
the Yukawa couplings of the Higgs field to the fermion
fields. In the case of quarks the responsible term of the
Lagrangian reads
\begin{equation}
 \quad - \!\sum_{j,k=1}^3 Y^u_{jk} \ov u{}_j^L u_k^R (v+H) 
  \, -\! 
    \sum_{j,k=1}^3 Y^d_{jk} \ov d{}_j^L d_k^R (v+H) + \mbox{h.c.}
\qquad
\parbox{2cm}{\includegraphics[height=1.8cm]{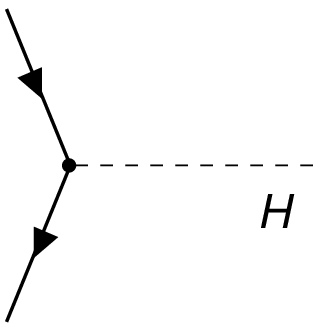}}. 
\label{yuk} 
\end{equation}
Here $H$ denotes the field of the yet-to-be-discovered physical Higgs
boson and $v=174\gev$ is the corresponding vacuum expectation value. The
indices $j$ and $k$ label the generations and $L$ and $R$ refer to the
chirality of the quark fields.  The Yukawa couplings for up-type and
down-type quarks are $3\times 3$ matrices in flavour space, denoted by
$Y^u$ and $Y^d$, respectively. Eq.~(\ref{yuk}) entails the mass matrices
\begin{equation}
 m^u= Y^u v \qquad \mbox{and}\qquad 
 m^d= Y^d v.  
\label{mama} 
\end{equation}
The diagonalisations of $m^u$ and $m^d$ involve four unitary rotations
in flavour space, one each for $u^L$, $u^R$, $d^L$, and $d^R$. 
Since the left-handed fields $u^L_k$ and $d^L_k$, which were originally
members of a common SU(2) doublet, undergo different rotations, the
electroweak SU(2) symmetry is no more manifest in the physical basis in
which mass matrices are diagonal. The mismatch between the rotations of
the left-handed fields defines the unitary Cabibbo-Kobayashi-Maskawa
(CKM) matrix \cite{c,km},
\begin{equation}
V = \left( \begin{array}{ccc}
                V_{ud} & V_{us} & V_{ub} \\
                V_{cd} & V_{cs} & V_{cb} \\
                V_{td} & V_{ts} & V_{tb}
        \end{array} \right). \label{defv}
\end{equation}
The CKM elements occur in the couplings of the $W$ boson, because the
$W$ e.g.\ couples the $\ov c^L$ field to the linear combination $V_{cd}
d^L+V_{cs} s^L + V_{cb} b^L$ as a consequence of the unitary rotations
in flavour space. $V$ can be parametrised in terms of three angles and
one phase, the CP-violating Kobayashi-Maskawa phase $\gamma$
\cite{km}. 
With a history of more than 50 years, research in quark flavour has been
essential for the construction of the Standard Model, having guided us
to phenomena which were ``new physics'' at their time: Highlights were
the breakdown of the discrete symmetries P \cite{ly} and CP
\cite{ccft,km}, the prediction of the charm quark \cite{gim} and its
mass \cite{gl}, and a heavy top quark predicted from the size of \bbmd\
\cite{argus}.  In the decade behind us the asymmetric B factories BELLE
and BaBar have consolidated the CKM picture of quark flavour
physics. With the advent of the LHC era, the focus of flavour physics
has shifted from CKM metrology to physics beyond the Standard Model. In
the Standard Model flavour-changing neutral current (FCNC) processes
(such as meson-antimeson mixing, $B\to X_s \gamma$ or $K\to \pi \ov\nu
\nu$) are forbidden at tree-level and only occur through highly
suppressed one-loop diagrams. FCNC processes are therefore excellent
probes of new physics.  This is a strong rationale to complement the
high-$p_T$ physics programs at ATLAS and CMS with precision flavour
physics at LHCb, NA62, BELLE-II, Super-B, BES-III, J-PARC and the future
intense proton source Project X at Fermilab.

With the discovery of neutrino flavour oscillations, the much younger
field of lepton flavour physics has emerged. The Standard Model in its
original formulation \cite{gsw} lacks a right-handed neutrino field and
can neither accommodate neutrino masses nor neutrino oscillations. The
simplest remedy for this is the introduction of a dimension-5 Yukawa
term composed of two lepton doublets $L=(\nu_{\ell}^ L,\ell^L)$ and two
Higgs doublets
leading to Majorana masses for the neutrinos and generating the desired
lepton flavour mixing. Alternatively one can mimic the quark sector by
introducing right-handed neutrino fields (and imposing $B-L$, the
difference between baryon and lepton numbers, as an exact symmetry).
With both variants FCNC transitions among charged leptons (such as
$\mu\to e\gamma$) are unobservably small, so that any observation of
such a process will imply the existence of further new
particles. Charged-lepton FCNC decays are currently searched for in the
dedicated MEG experiment (studying $\mu\to e\gamma$), in B factory data
(on e.g.\ $\tau \to \mu\gamma$) and at three of the four major LHC
experiments (searching e.g.\ for $\tau\to\mu\ov \mu\mu$).

In the following section I briefly review recent theoretical progress on
the Standard-Model predictions for FCNC processes. Subsequently I
discuss new developments in flavour physics beyond the Standard Model.
I limit myself to supersymmetric theories, which reflects my personal
research interests. For a recent broader overview, which also 
covers extra dimensions and Little-Higgs models, 
see ref.~\cite{b}. Exhaustive studies of the flavour sector 
in a four-generation  Standard Model can be found in refs.~\cite{fourgen}.

\section{Standard Model} 
\begin{figure}
\includegraphics[scale=.6, angle=0,
                 clip=false]{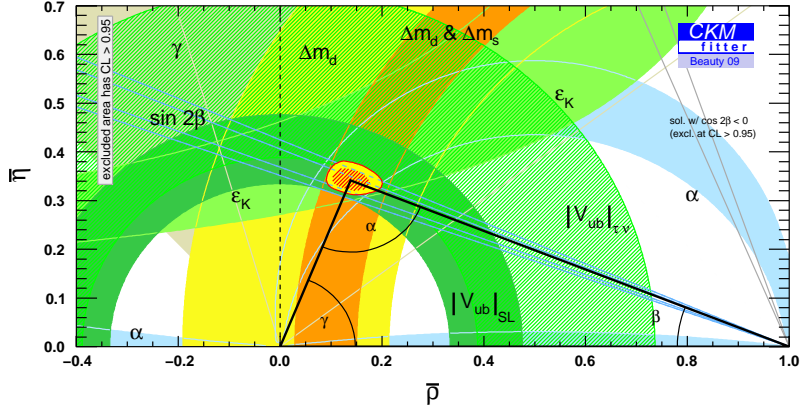} 
\caption{Experimental constraints on the unitarity triangle, 
 from ref.~\cite{ckmf}. }\label{fig:ut}
\end{figure}

The standard unitarity triangle (UT) is a triangle with unit baseline 
and apex $(\ov\rho,\ov\eta)$, which is defined through
\begin{equation}
\ov \rho + i \ov \eta \equiv - \frac{V_{ub}^* V_{ud}}{V_{cb}^* V_{cd}}
 \label{defre} .
\end{equation}
The two non-trivial sides of the triangle
are 
$  R_u \equiv \sqrt{\ov \rho^2 + \ov \eta^2 }$ and 
$ R_t \equiv  \sqrt{(1-\ov \rho)^2 + \ov \eta^2} $.
The triangle's three angles
\begin{equation}
    \alpha = \arg\left[ - \frac{V_{td}V_{tb}^*}{V_{ud}V_{ub}^*}\right],
    \qquad
    \beta  = \arg\left[ - \frac{V_{cd}V_{cb}^*}{V_{td}V_{tb}^*}\right],
    \qquad
    \gamma = \arg\left[ - \frac{V_{ud}V_{ub}^*}{V_{cd}V_{cb}^*}\right].
    \label{eq:beta}
\end{equation}
are associated with CP-violating quantities. Measurements of
flavour-changing quantities imply constraints on $(\ov\rho,\ov\eta)$.
Last year's global analysis of the UT performed by the CKMfitter
collaboration is shown in \fig{fig:ut}. For the results of the UTFit
collaboration, which uses a different statistical approach see 
ref.~\cite{utf}. 
The figure shows the consistency of the various measurements, which
single out the small yellow area as the allowed region for the apex of
the triangle. Clearly, the CKM mechanism is the dominant source of
flavour violation in the quark sector. 

From the quantities entering the global UT analysis in \fig{fig:ut} 
the meson-antimeson mixing amplitudes are the ones most 
sensitive to generic new physics. 
While the extraction of the UT angle $\beta$ from the CP phase in \bbmd\
is theoretically very clean, all other quantities related to
meson-antimeson mixing are plagued by theoretical uncertainties. Namely,
the uncertainties in the mass differences $\Delta m_q$ and $\Delta m_s$
of the two \bbm\ complexes and in $\epsilon_K$, which quantifies CP
violation in \kkm, completely dominate over the irrelevantly small
experimental errors. Note that $\Delta m_s$ is practically independent
of $\ov\rho$ and $\ov\eta$ and is only useful for the UT fit because the
ratio $\Delta m_d/\Delta m_s$ has a smaller uncertainty than $\Delta
m_d$. The \kkm\ mixing amplitude $M_{12}$ 
involves the matrix element $\bra{K^0} H^{\Delta S=2} \ket{\ov K{}^0}$
of the $\Delta S=2$ hamiltonian $ H^{\Delta S=2}$ \cite{bjw}. $H^{\Delta
  S=2}$ is proportional to the four-quark operator $\ov{d}{}^L
\gamma_{\nu} s{}^L \, \ov{d}{}^L \gamma^{\nu} s^L$ with the relevant
matrix element
\begin{equation}
   \bra{K^0}  \ov{d}{}^L \gamma_{\nu} s{}^L \, \ov{d}{}^L \gamma^{\nu} s^L  
   (\mu) \ket{\ov K{}^0} =   \frac{2}{3} M_{K}^2 \, f_K^2 \,
  \frac{\widehat B_K}{ b(\mu )} . \label{melk}
\end{equation}
This equation merely defines the parameter $\widehat B_K$ which is
commonly used to parametrise the matrix element of interest. In
\eq{melk} $M_K=497.6\mev$ and $f_K =160\mev$ are mass and decay constant
of the neutral Kaon and $b_K(\mu)$ is introduced to render $\widehat
B_K$ independent of the unphysical renormalisation scale $\mu$ and the
renormalisation scheme chosen for the definition of the 
operator $\ov{d}{}^L \gamma_{\nu} s{}^L \, \ov{d}{}^L \gamma^{\nu} s^L
(\mu)$. In the commonly used $\ov{\rm MS}$ scheme one has
$b_K(\mu=1\,\gev)=1.24\pm 0.02$. 
The matrix element in \eq{melk} must be calculated with lattice gauge
theory. A new computation by Aubin, Laiho and Van de Water finds
\cite{alv}
\begin{equation}
   \widehat B_K = 0.724(8)(29)  
\end{equation}
This result is in good agreement with the 2007 result of the RBC and
UKQCD collaborations, $ \widehat B_K = 0.720(13)(37)$
\cite{Cohen:2007zz}. In view of the superb experimental precision in
$|\epsilon_K|= (2.23 \pm 0.01) \times 10^{-3}$ further progress on
$\widehat B_K$ is certainly highly desirable. The increasing precision
in $\widehat B_K $ has also stimulated more precise analyses of other
ingredients of $M_{12}$. Recently a reanalysis of the long-distance
contribution to $\imag M_{12}$ has resulted in an upward shift of 2\%\
in $\epsilon_K$ \cite{bgi}. A similar contribution constituting the
element $\Gamma_{12}$ of the decay matrix, affects $\epsilon_K$ at the
few-percent level \cite{run2,bg}.
 
In the case of \bbm\ all long-distance contributions are highly
GIM-suppressed and only the local contribution from the box diagram with
internal top quarks and $W$ bosons matters. The two mass eigenstates of
the neutral \bbq\ system differ in their masses and widths.  The mass
difference $\dm_q$, $q=d,s$, which equals the \bbq\ oscillation
frequency, is given by $\dm_q\simeq 2| M_{12}^q| = 2| \bra{B_q}
H^{\Delta B=2} \ket{\ov B_q}|$.  Lattice calculations are needed to
compute $f_{B_q}^2 \widehat B_{B_q}$, which is defined in analogy to
\eq{melk}. Here I focus on the ratio $\dm_d/\dm_s$ yielding the orange
(medium gray) annulus centred around $(\ov\rho,\ov \eta)=(1,0)$ in
\fig{fig:ut}. This ratio involves the hadronic quantity
\begin{eqnarray}
 \xi & =&
   \frac{f_{B_s} \sqrt{\widehat{B}_{B_s}}}{
              f_{B_d} \sqrt{\widehat{B}_{B_d}}}
      \;=\; 1.23 \pm 0.04 .
\label{xi}
\end{eqnarray}
The numerical value in \eq{xi} is my bold average of the values
summarised by Aubin at the \emph{Lattice'09} conference \cite{a}. 
With this number and the measured values 
$\dm_{B_d} = ( 0.507 \pm 0.005 ) \,
  \mbox{ps}^{-1}$ \cite{pdg} and 
$\dm_{B_s} = ( 17.77\pm 0.10 \pm 0.07 ) \, \mbox{ps}^{-1}$ \cite{cdf}
one finds
\begin{eqnarray}
\lt| \frac{V_{td}}{V_{ts}} \rt| &=&
   \sqrt{\frac{\dm_{B_d}}{\dm_{B_s}}} \,
   \sqrt{\frac{M_{B_s}}{M_{B_d}}} \, \xi = 
   0.210 \pm 0.007.    
   \label{vdmds}
\end{eqnarray}
With $|V_{td}/V_{ts}| = 0.228 \, R_t$ one finds $R_t=0.92\pm 0.03$ for the 
side of the UT opposite to $\gamma$. For a pedagogical introduction into 
meson-antimeson mixing and CKM phenomenology cf.\ ref.~\cite{n}.

\section{Beyond the Standard Model}
\subsection{Phenomenology of new physics in
  $B\!-\!\ov{B}{}\,$\ mixing }
\begin{figure}
\includegraphics[width=0.48\textwidth]{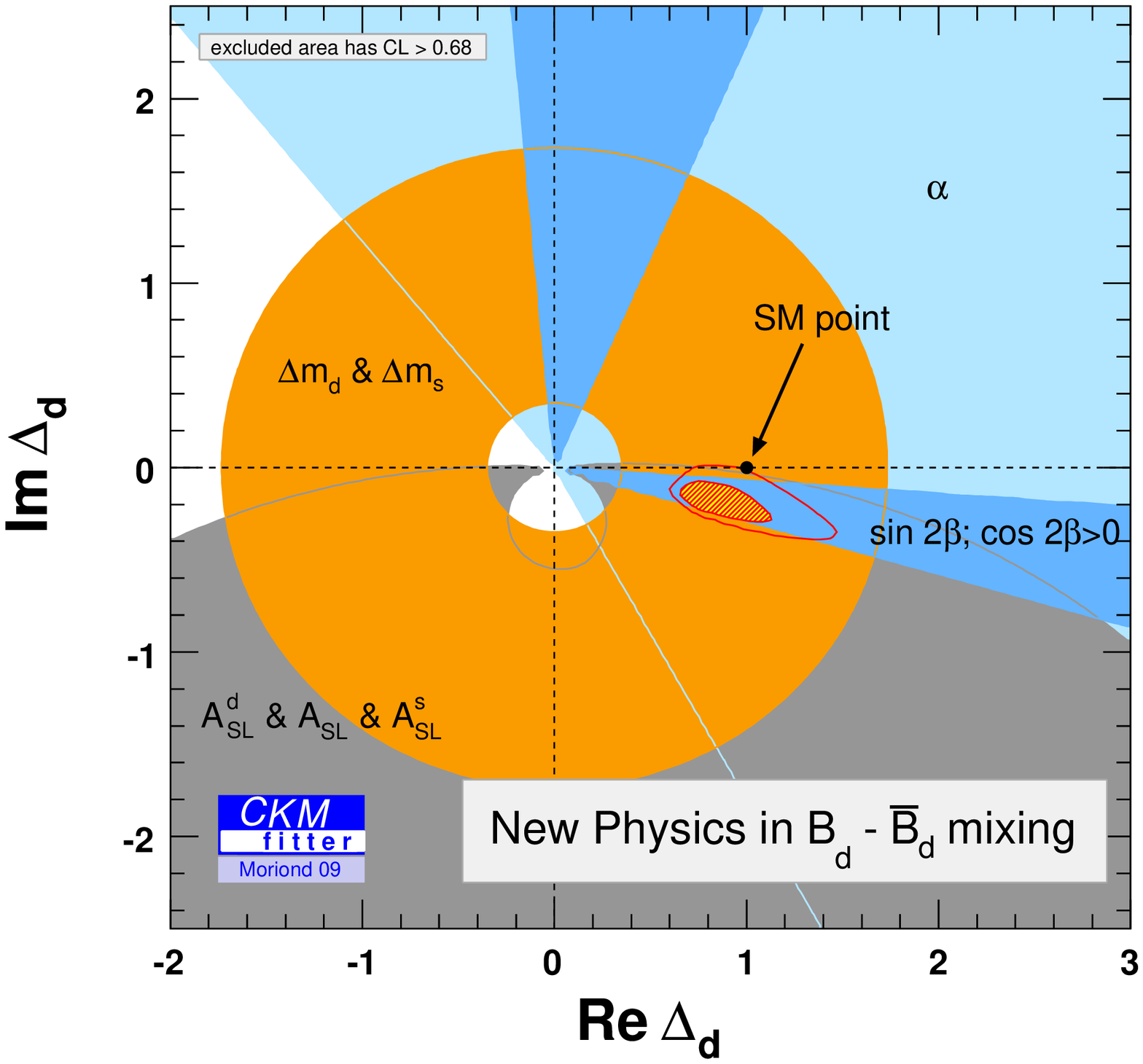} 
\hfill
\includegraphics[width=0.48\textwidth]{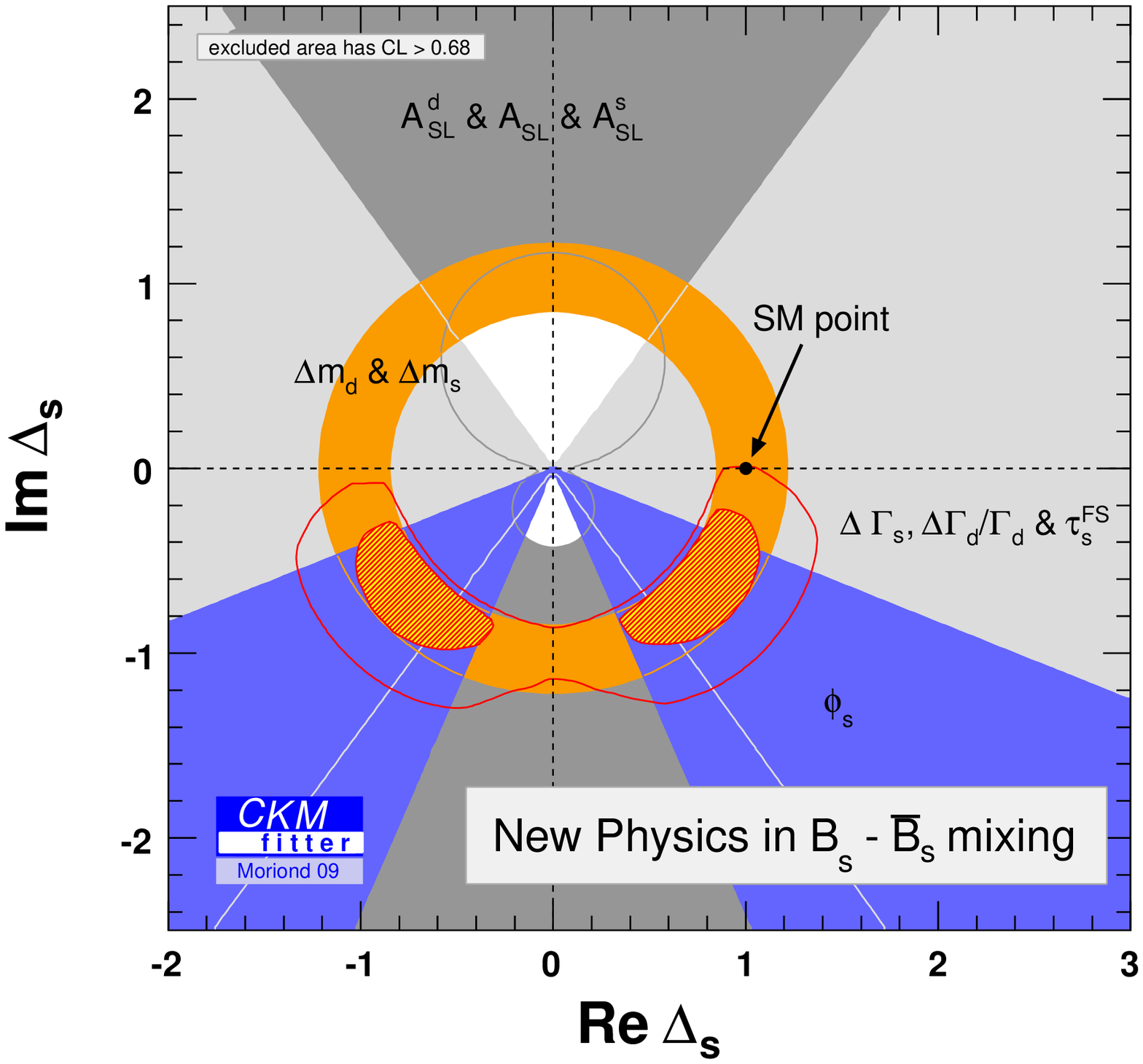} 
\caption{Complex $\Delta_d$ and $\Delta_s$ planes, plots taken from the
  web site in ref.~\cite{ckmf}. (See ref.~\cite{ckmfb} for details of
  the analysis.) Similar analyses by the UTfit collaboration can be
  found in refs.~\cite{utf,utfb}.}\label{fig:bds}
\end{figure}

The plot of the UT in \fig{fig:ut} is not the best way to show possible
deviations from the Standard Model, because it conceals certain
correlations between different quantities. In the LHC era we will more
often see plots of quantities which directly quantify the size of new
physics contributions. In the case of meson-antimeson mixing new physics
can be parametrised model-independently by a single complex parameter
\cite{ln}.  For \bbmq, $q=d,s$, one defines
\begin{equation}
   \Delta_q = \frac{M_{12}^q}{M_{12}^{q,\rm SM}}. 
   \label{defdel}
\end{equation}
The CKMfitter collaboration has found that the Standard-Model point
$\Delta_d=1$ is ruled out at 95\%\ CL (left plot in \fig{fig:bds}), if
all other quantities entering the global UT analysis are assumed free of
new physics contributions. This discrepancy is largely driven by
$B(B^+\to\tau^+\nu)$ and, if interpreted in terms of new physics, may
well indicate non-standard physics in quantities  other than \bbmd. A
tension on the global UT fit was also noted by Lunghi and Soni \cite{ls}
and by Buras and Guadagnoli \cite{bg}. The situation is much simpler in
the case of \bbms, which shows a deviation from the Standard Model
expectation of similar size (right plot in \fig{fig:bds}). 
The allowed region for $\Delta_s$ is essentially independent from input
other than the \bbms\ amplitude $M_{12}^s$. The quantities entering the
analysis are primarily $\dm_s$, the width difference $\dg_s$
\cite{bbgln,ln}, the time-dependent angular distribution in $B_s\to
J/\psi \phi$ (with access to the mixing-induced CP asymmetry $A_{\rm
  CP}^{\rm mix}(B_s\to J/\psi \phi)$ if the $B_s$ flavour is tagged),
and the CP asymmetry in flavour-specific decays $a_{\rm fs}^s$
\cite{bbgln,ln}.  The first global analysis of these quantities, which
used improved Standard-Model predictions, was performed in 2006
\cite{ln} showing a 2$\sigma$ deviation from the Standard-Model value
$\Delta_s=1$. At the time of this talk the discrepancy from the combined
D\O\ and CDF data on $B_s\to J/\psi \phi$ alone was between 2.0$\sigma$
and 2.3$\sigma$, depending on details of the statistical analysis
\cite{comb}.  After this conference the discrepancy in $a_{\rm fs}^s$
has increased due to a new D\O\ measurement of the dimuon asymmetry in a
mixed $B_d,B_s$ data sample \cite{d0dimuon}.  On the other hand, new CDF
data on $A_{\rm CP}^{\rm mix}(B_s\to J/\psi \phi)$ have pulled the
result towards the Standard Model \cite{newcdf}.  Still all measurements
favour $\arg{\Delta_s}<0$.

\subsection{Supersymmetry with large $\tan\beta$}
\begin{figure}
\centering
\includegraphics[width=6.5cm]{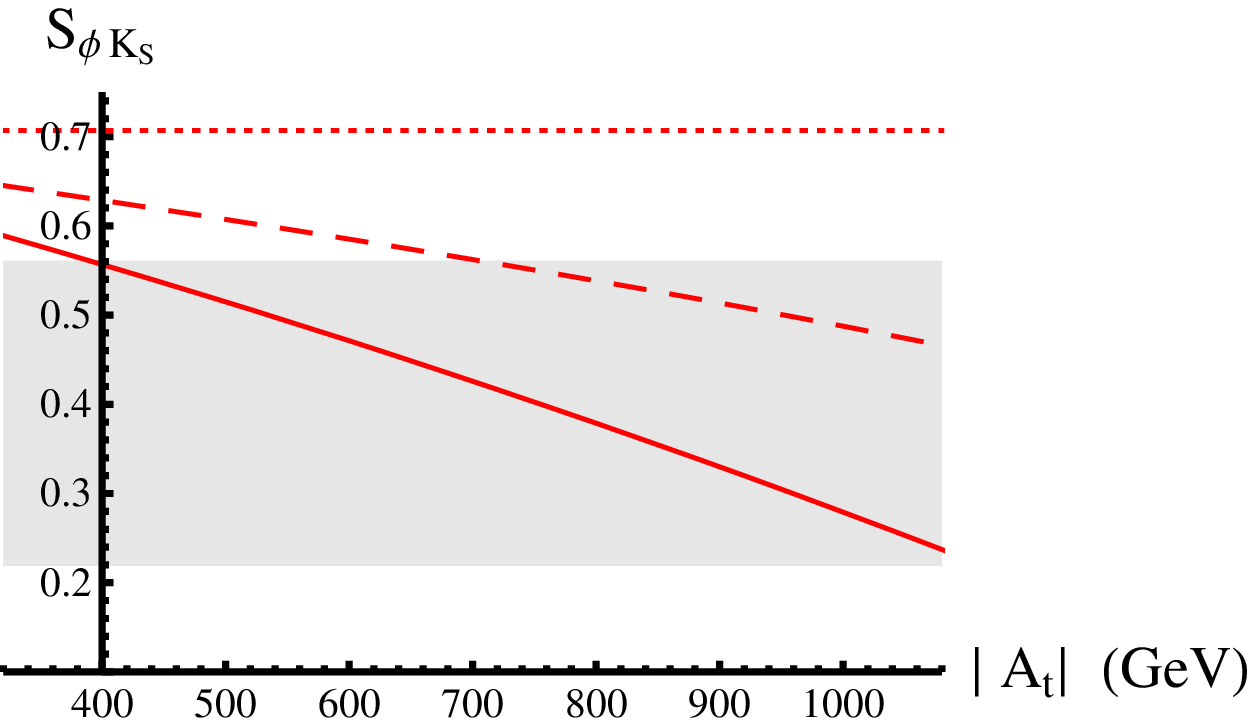}~~~~ 
\includegraphics[width=3.3cm]{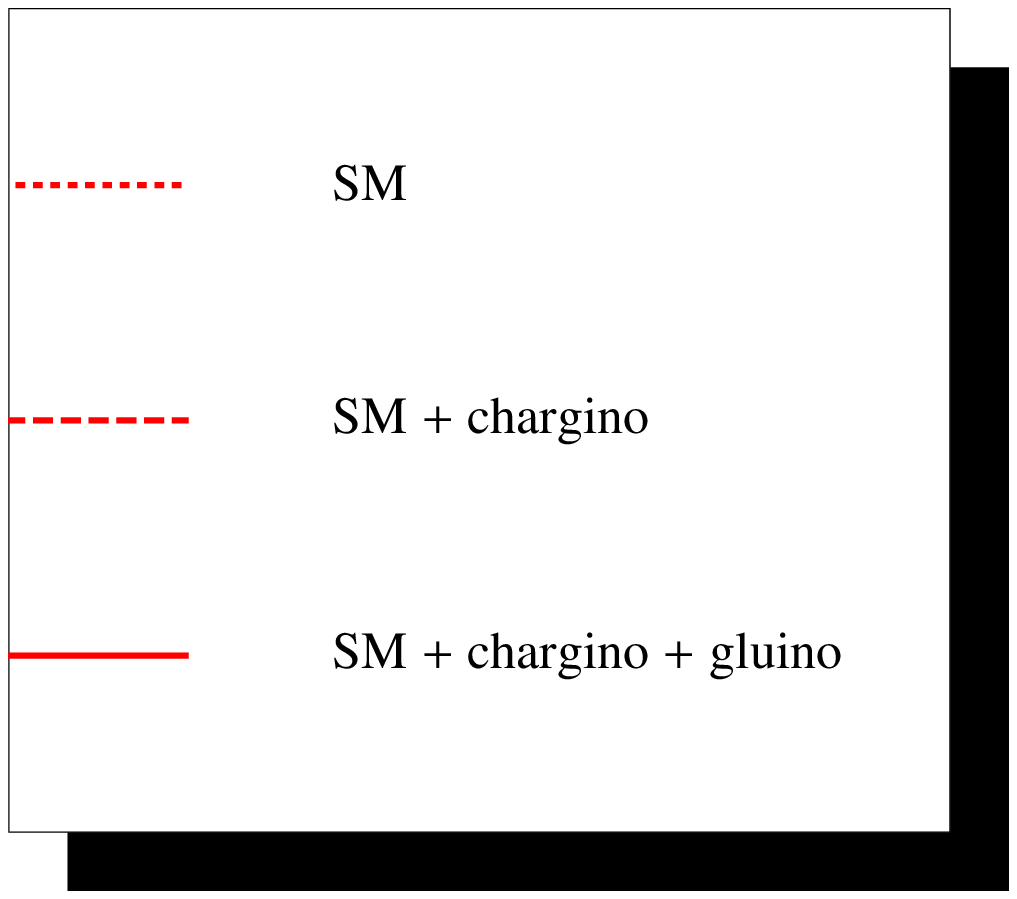} 
\caption{$S_{\phi K_S}$ as a function of $|A_t|$ for a parameter point
  compatible with other experimental constraints (see ref.~\cite{hns}
  for details). Solid: full result including the new contribution, 
  dashed: SM plus one-loop chargino diagram,  dotted: 
  SM value. Shaded area:
  experimental $1\sigma$ range.}\label{fig:spp}
\end{figure}

Extensions of the Standard Model typically come with new sources of
flavour violation, beyond the Yukawa couplings in \eq{yuk}.  
In the Minimal Supersymmetric Standard Model (MSSM) the soft
supersymmetry breaking terms a priori possess a flavour structures which
is unrelated to $Y^u$ and $Y^d$. To avoid excessive FCNCs violating
experimental bounds the MSSM is often supplemented with the assumption
of Minimal Flavour Violation (MFV), which amounts to a flavour-blind
supersymmetry-breaking sector. In the MFV-MSSM supersymmetric FCNC
transitions are typically smaller than the error bars of today's
experiments, unless the parameter $\tan\beta$ is large. Probing values
around $\tan\beta=60$ tests the unification of top and bottom Yukawa
couplings. Importantly, loop suppression factors can be offset by a
factor of $\tan\beta$ and may yield contributions of order one, with
most spectacular effects in $B(B_s\to \mu^+\mu^-)$ \cite{bk}. The
$\tan\beta$-enhanced loop corrections must be summed to all orders in
perturbation theory. In the limit that the masses of the SUSY particles
in the loop are heavier than the electroweak vev and the masses of the
five Higgs bosons, $M_{\rm SUSY} \gg v, M_{A^0},M_{H^+}\ldots$, one can
achieve this resummation easily: After integrating out the heavy SUSY
particles one obtains an effective two-Higgs doublet model with novel
loop-induced couplings \cite{hrs}.  
In supersymmetric theories, however, it is natural that $M_{\rm SUSY}$
is not much different from $v$ and further $M_{\rm SUSY} \gg M_{A^0}$
involves an unnatural fine-tuning in the Higgs sector.
Phenomenologically, large-$\tan\beta$ scenarios comply with the
experimental bound from $B(B_s\to \mu^+\mu^-)$ more easily if $M_{A^0}$
is large, which may easily conflict with $M_{\rm SUSY} \gg M_{A^0}$. To
derive resummation formulae valid for arbitrary values of $M_{\rm SUSY}$
one cannot resort to the method of an effective field theory.  Instead
one should work strictly diagrammatically in the full MSSM to identify
$\tan\beta$-enhanced corrections. This procedure requires full control
of the renormalisation scheme: The analytical results for the resummed
expressions differ for different schemes and not all renormalisation
schemes permit an analytic solution to the resummation problem. The
diagrammatic resummation has been obtained for the flavour-diagonal case
in ref.~\cite{cgnw} and recently for flavour-changing interactions in
ref.~\cite{hns}. This opens the possibility to study
$\tan\beta$-enhanced corrections also to supersymmetric loop processes
which decouple for $M_{\rm SUSY} \gg v$ and to collider processes
involving supersymmetric particles. In ref.~\cite{hns} a novel large
effect, which does not involve Higgs bosons, in the Wilson coefficient
$C_8$ has been found, with interesting implications for the
mixing-induced CP asymmetry $S_{\phi K_S}$ in $B_d \rightarrow \phi K_S$
(see \fig{fig:spp}).

\subsection{Radiative flavour violation\label{sect:rfv}}
A symmetry-based definition of MFV starts from the observation that the
MSSM sector is invariant under arbitrary unitary rotations of the
(s)quark multiplets in flavour space.  This $[$U(3)$]^3$ flavour
symmetry ($[$U(3)$]^5$ if (s)leptons are included) is broken by the
Yukawa couplings, and MFV can be defined through the postulate that the
Yukawa couplings are the only spurion fields breaking the $[$U(3)$]^3$
flavour symmetry \cite{D'Ambrosio:2002ex}. Interestingly, there is a
viable alternative to MFV to solve the supersymmetric flavour problem:
We may start with a Yukawa sector in which all Yukawa couplings of the
first and second generation are zero. That is, the MSSM superpotential
possesses an \emph{exact} $[U(2)]^3 \times U(1)$ symmetry. Then we
postulate that the trilinear SUSY breaking terms $A^u_{ij}$ and
$A^d_{ij}$ are the spurion fields breaking this symmetry. The observed
off-diagonal CKM elements and the light quark masses are generated
radiatively through squark-gluino loops, explaining their smallness in a
natural way. In ref.~\cite{Crivellin:2008mq} it has been found that this
setup of Radiative Flavour Violation (RFV) complies with all FCNC
bounds, if the squark masses are larger than roughly $500\gev$. By
contrast, the bilinear SUSY breaking terms cannot be the spurion fields
breaking $[U(2)]^3 \times U(1)$ without violating the constraints from
FCNC processes. The idea that SUSY breaking could be the origin of
flavour violation is not new \cite{bw,Borzumati:1997bd}, remarkably
the absence of tree-level light-fermion Yukawa couplings substantially
alleviates the supersymmetric CP problem associated with electric dipole
moments \cite{Borzumati:1997bd}.  

The finding that loop contributions involving $A^q_{ij}$, $q=u,d$, can
be large has also consequences for the generic MSSM: In FCNC analyses
aiming at constraints on flavour-violating SUSY-breaking terms one must
include chirally enhanced higher-order corrections involving $A^q_{ij}$
and, if $\tan\beta$ is large, also corrections with bilinear
SUSY-breaking terms \cite{Crivellin:2009ar}. The trilinear terms further
imply important loop corrections to quark and lepton masses
\cite{Gabbiani:1996} and can induce right-handed $W$
couplings \cite{Crivellin:2009sd}.

\subsection{MSSM with GUT constraints}
In grand unified theories (GUTs) quarks and leptons are combined into
symmetry multiplets. As a consequence, it may be possible to see
imprints of lepton mixing in the quark sector and vice versa. In
particular, the large atmospheric neutrino mixing angle may influence
$b\to s$ transitions through the mixing of right-handed $\tilde b$ and
$\tilde s$ squarks \cite{m}. Yet the usual small dimension-4 Yukawa
interactions of the first two generations are sensitive to corrections
from dimension-5 terms which are suppressed by $M_{\rm GUT}/M_{\rm
  Planck}$ \cite{Ellis:1979fg}. These contributions are welcome to fix
the unification of the Yukawa couplings, but may come with an arbitrary
flavour structure, spoiling the predictiveness of the quark-lepton
flavour connection.  
SU(5) and SO(10) models with dimension-5 Yukawa couplings have been
studied in great detail \cite{Baek:2001kh}.  Phenomenologically one can
constrain the troublesome flavour misalignment using data on FCNC
transitions between the first two generations. Here I present a recent
SU(5) analysis exploiting the experimental bound on $B(\mu\to e \gamma)
$ \cite{Girrbach:2009uy}. At the GUT scale the Yukawa matrices
for down-type quarks, $Y_d$ and $Y_l$, read
\begin{equation}
   Y_d =
      Y_{\rm GUT} + k_d\, \frac{\sigma}{M_{\rm Planck}} Y_\sigma \;, \qquad
   Y_l^\top
   = Y_{\rm GUT} + k_e\, \frac{\sigma}{M_{\rm Planck}} Y_\sigma .
\end{equation}
Here $Y_{\rm GUT}$ is the unified dimension-4 Yukawa matrix,
$\sigma=\mathcal{O}\left(M_{\rm GUT}\right)$ is a linear combination of
Higgs vevs and the prefactors $k_d$ and $k_e $ differ from each other
due to GUT breaking.  If the universality condition $A_l = A_d = a_0
Y_{\rm GUT}$ is invoked at the GUT scale, any misalignment between $
Y_{\rm GUT}$ and $ Y_\sigma$ will lead to a non-MFV low-energy theory,
because $ A_l \propto\hspace{-3mm}/\hspace{2mm} Y_l$ and $ A_d
\propto\hspace{-3mm}/\hspace{2mm} Y_d$.
We may parametrise this effect as
\begin{equation}
  A_l  \simeq A_{0}\lt(
  \begin{array}{ccc}
    \cos\theta & -\sin\theta & 0 \cr \sin\theta & \cos\theta & 0 \cr 0 &
    0 & 1
  \end{array}\rt)
  Y_{l} .
\end{equation}
Now the experimental upper bound on $B(\mu\to e \gamma) $ determines the
maximally allowed $|\theta|$ as a function of $A_0$. In
ref.~\cite{Girrbach:2009uy} it is found that $|\theta|$ can hardly
exceed 10 degrees once $|A_0|$ exceeds 50$\,\gev$. An analysis in the
quark sector (studying SO(10) models \cite{m}) finds similar strong
constraints from $\epsilon_K$ \cite{Trine:2009ns}. As a consequence, the
dimension-5 terms can barely spoil the GUT prediction derived from
the dimension-4 relation $ Y_d = Y_l^\top = Y_{\rm GUT}$, unless $|A_0|$
is small.  This result may indicate that dimension-4 and dimension-5
Yukawa couplings are governed by the same flavour symmetries. 

\section*{Acknowledgements}
I thank the organisers for the invitation to this wonderful
conference and for financial support.  The presented work is supported by
the DFG Research Unit SFB--TR 9, by BMBF grant no.~05H09VKF and by the
EU Contract No.~MRTN-CT-2006-035482, \lq\lq FLAVIAnet''.

\end{document}